\documentclass[a4paper,11pt]{article}
\usepackage{pos}
\begin{document}
\title{The LHCspin project}
\author*[a]{P. Di Nezza}
\author[b,c]{V. Carassiti}
\author[b,c]{G. Ciullo}
\author[b,c]{P. Lenisa}
\author[b,c]{L.L. Pappalardo}
\author[a]{M. Santimaria}
\author[d]{E. Steffens}
\author[e]{G. Tagliente}
\affiliation[a]{Istituto Nazionale di Fisica Nucleare, Laboratori Nazionali di Frascati, 00044
Frascati, Italy}
\affiliation[b,c]{Istituto Nazionale di Fisica Nucleare, Sezione di Ferrara, 44122 Ferrara, Italy, Dipartimento di Fisica e Scienze della Terra,
Università di Ferrara, 44122 Ferrara, Italy}
\affiliation[d]{Physikalisches Institut, Universität Erlangen-Nürnberg, 91058 Erlangen, Germany}
\affiliation[e]{Istituto Nazionale di Fisica Nucleare, Sezione di Bari, 70125 Bari, Italy}
\emailAdd{Pasquale.DiNezza@lnf.infn.it}
\abstract{
LHCspin aims to upgrade the recently installed unpolarized gas target (SMOG2) in front of the LHCb spectrometer to a polarised one. This task requires, in the next few years, innovative solutions and cutting-edge technologies, and will allow the exploration of a unique kinematic regime and new reaction processes. 
With the instrumentation of the proposed target system, LHCb will become the first experiment delivering simultaneously unpolarized beam-beam at $\sqrt{s}$=14 TeV, and unpolarized and polarized beam-target collisions at $\sqrt{s_{NN}}\sim$100 GeV. LHCspin could open new physics frontiers exploiting the potential of the most powerful collider and one of the most advanced detectors.}
\FullConference{PANIC 2021 \\ Lisbon \\
5-10 September 2021
}
\maketitle

\section{The scientific case}

The LHCb detector \cite{lhcb} is a general-purpose forward spectrometer specialised in detecting hadrons containing c and b quarks, fully instrumented in the 2 < $\eta$ < 5 region, and is the only detector able to collect data in both collider and fixed-target mode.

\begin{figure}[!h]
\centering
\includegraphics[width=10cm]{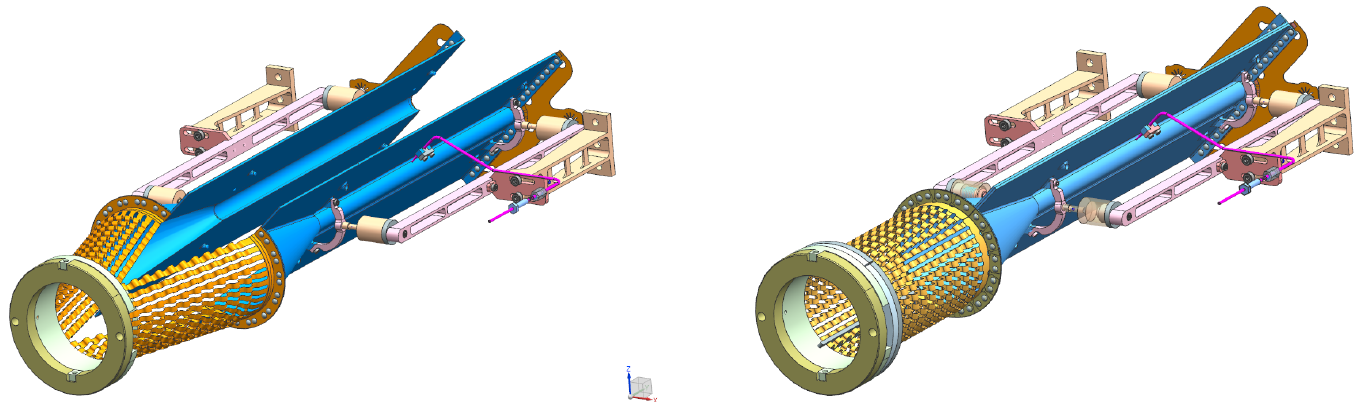}
\caption{\label{cell} The SMOG2 storage cell in the open (left) and closed (right) configuration.}
\end{figure}

The fixed-target physics program at LHCb is active since the installation of the SMOG (System for Measuring the Overlap with Gas) device \cite{smog}, and now, with the SMOG2 upgrade \cite{s2}, an openable gas storage cell, shown in Fig.~\ref{cell}, and an advanced Gas Feed System, a strong boost will come with the LHC Run 3. The target areal density will increase up to two orders of magnitude, depending on the injected gas species, and the data will be collected simultaneously for the beam-gas and beam-beam collisions.

The LHCspin project \cite{lhcspin} aims at extending the SMOG2 physics program in Run 4 and, with the installation of a polarised gas target, to bring spin physics at LHC for the first time, by exploiting the well suited LHCb detector.

The physics case of LHCspin covers three main areas: exploration of the wide physics potential offered by unpolarised gas targets, investigation of the nucleon spin, and heavy-ion collisions.

Similarly to SMOG2, LHCspin will allow the injection of several species of unpolarised gases giving excellent opportunities to investigate parton distribution functions (PDFs) in both nucleons and nuclei in the large-$x$ and intermediate $Q^2$ regime, and impact several fields of physics from QCD to astroparticle. Beside the collinear PDFs, polarised quark and gluon distributions can be probed by means of proton collisions on polarised hydrogen and deuterium, such as Generalised Parton Distributions (GPDs) and Transverse Momentum Dependent distribution functions (TMDs). In fact, there are several leading-twist distributions that can be probed with unpolarised and transversely polarised quarks and nucleons, giving independent information on the hadronic spin structure. 

\begin{figure}[!h]
\centering
\includegraphics[width=10cm]{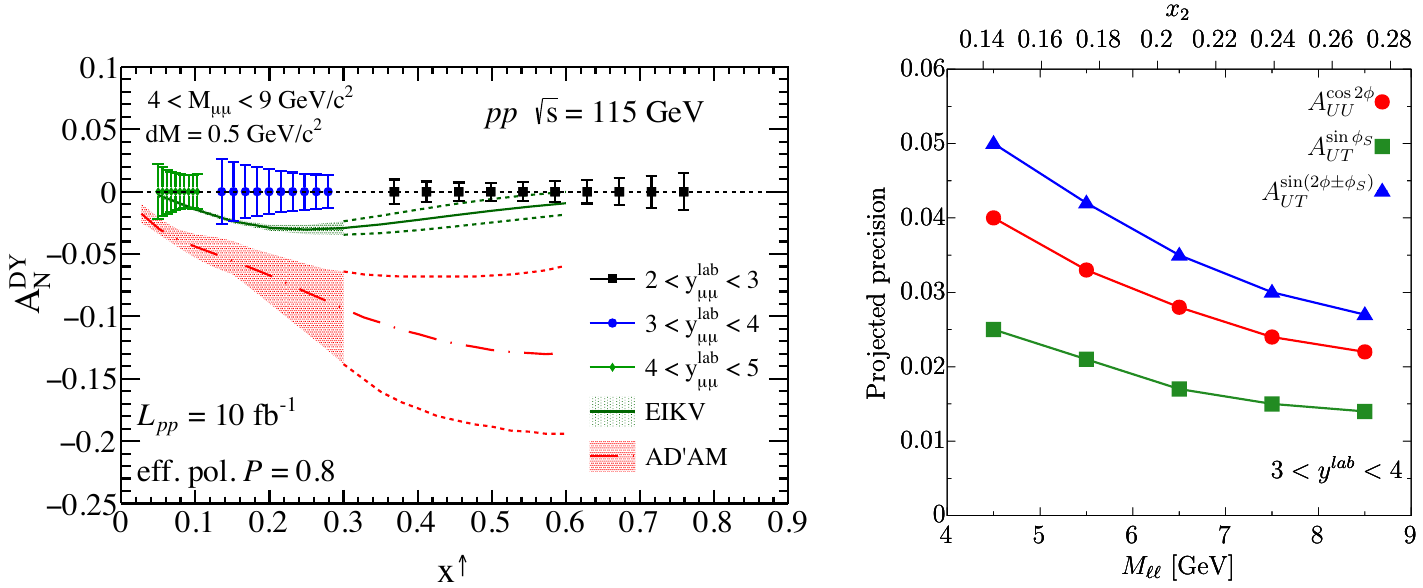}
\caption{\label{asym} Left: Measurements of Single Transverse Spin Asymmetry in DY events as a function of $x$ compared to two theoretical predictions \cite{dy}. Right: projected precision for some azimuthal asymmetry amplitudes with DY
data as a function of the dilepton invariant mass in the rapitidy range 3<$y^{lab}$<4 \cite{dy}.}
\end{figure}

To access the transverse motion of partons within a polarised nucleon, transverse single spin asymmetries have to be measured. For example, the polarised Drell-Yan (DY) channel probes the product of $f_1$ (unpolarised TMD) and $f_{1T}^{\perp}$ (Sivers function) in quarks and antiquarks in the low- and high-$x$ regimes, respectively. Projections for the uncertainty of such measurements are shown in Fig.~\ref{asym} (left) based on an integrated luminosity of 10 fb$^{-1}$. Being T-odd, it is theoretically established that the Sivers function changes sign in polarised DY with respect to semi-inclusive deep inelastic scattering \cite{sivers}. This fundamental QCD prediction can be verified by exploiting the large sample of DY data expected at LHCspin. In addition, isospin effects can be investigated by comparing $pH$ and $pD$ collisions. Several TMDs can be probed by evaluating the azimuthal asymmetries of the produced dilepton pair: projected precisions for three such asymmetries are shown in Fig.~\ref{asym} (right) for a specific rapidity range. 

\begin{figure}[!h]
\centering
\includegraphics[width=6cm]{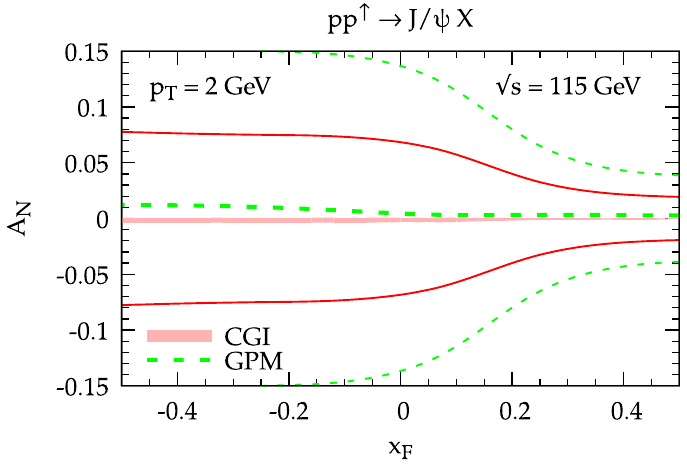}
\includegraphics[width=6cm]{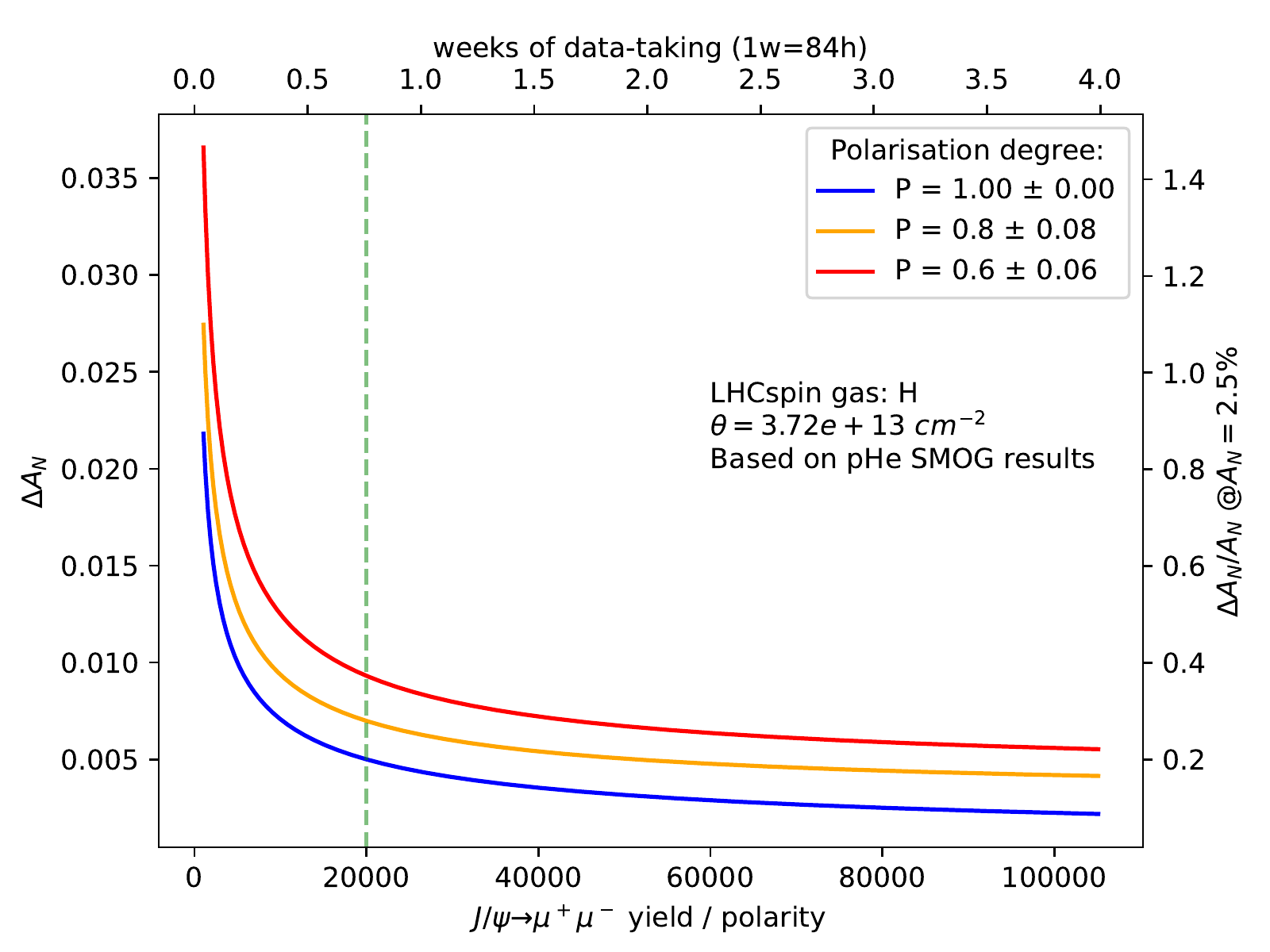}
\caption{\label{jpth} Left: theoretical predictions for $A_N$ in inclusive $J/\Psi$ production \cite{the}. Right: accuracy reacheable as a function of the collected number of $J/\Psi$, the data taking time and for different values of the target polarization.}
\end{figure}

Heavy flavour states will be one of the strength points of LHCspin. Being mainly produced via gluon fusion at LHC, quarkonia and open heavy flavour states will allow to probe the unknown gluon Sivers function via inclusive production of $J/\Psi$ or $D^0$ but also with several unique states like $\eta_c$, $\chi_c$, $\chi_b$
or $J/\Psi ~J/\Psi$. Fig.~\ref{jpth} (left) shows two predictions, based on the analysis performance of \cite{phe}, for the asymmetry on $J/\Psi$ events: 5 - 10 \% asymmetries are expected in the $x_F$ < 0 region, where the LHCspin sensitivity is the highest. Fig.~\ref{jpth} (right) shows the reachable precision as a function of the number of $J/\Psi$ particles that can be collected in just few days of data taking, all for three different values of the target polarisation. Moreover, heavy flavour states can be exploited as well to probe the gluon-induced asymmetries $h_1^{\perp g}$ (Boer-Mulders) and $f_1^g$, which are both experimentally unconstrained.

The long list of potential measurements based on the use of the LHC proton beam will be further enriched by using the lead beam.
For exampe, the Quark Gluon Plasma phase diagram exploration can be performed with a rapidity scan at a centre of mass energy, which is in-between those reached at RHIC and SPS. Moreover, LHCspin gives 
the unique possibility to merge the LHC heavy-ion program with spin physics, allowing, for the first time, to measure polarized Pb-p$^{\uparrow\downarrow}$ and Pb-d$^{\uparrow\downarrow}$ collisions at $\sqrt{s_{NN}} \sim$ 72 GeV. Among the others, flow measurements will greatly benefit from the excellent identification performance of LHCb on charged and neutral light hadrons.
The dynamics of small systems is an interesting topic joining heavy-ion collisions and spin physics where, in the spin 1 deuteron nucleus, the nucleon matter distribution is prolate for $j_3$ = $\pm$1 and oblate for $j_3$ = 0, where $j_3$ is the projection of the spin along the polarisation axis. In ultra-relativistic lead ion
collisions on transversely polarised deuteron, the deformation of the target deuteron can influence the orientation of the fireball in the transverse plane, quantified by the ellipticity. The measurement proposed in \cite{hi} can easily be performed at LHCspin on minimum bias events thanks to the high-intensity LHC beam.

\section{The experimental setup}

\begin{figure}[!h]
\centering
\includegraphics[width=9.7cm]{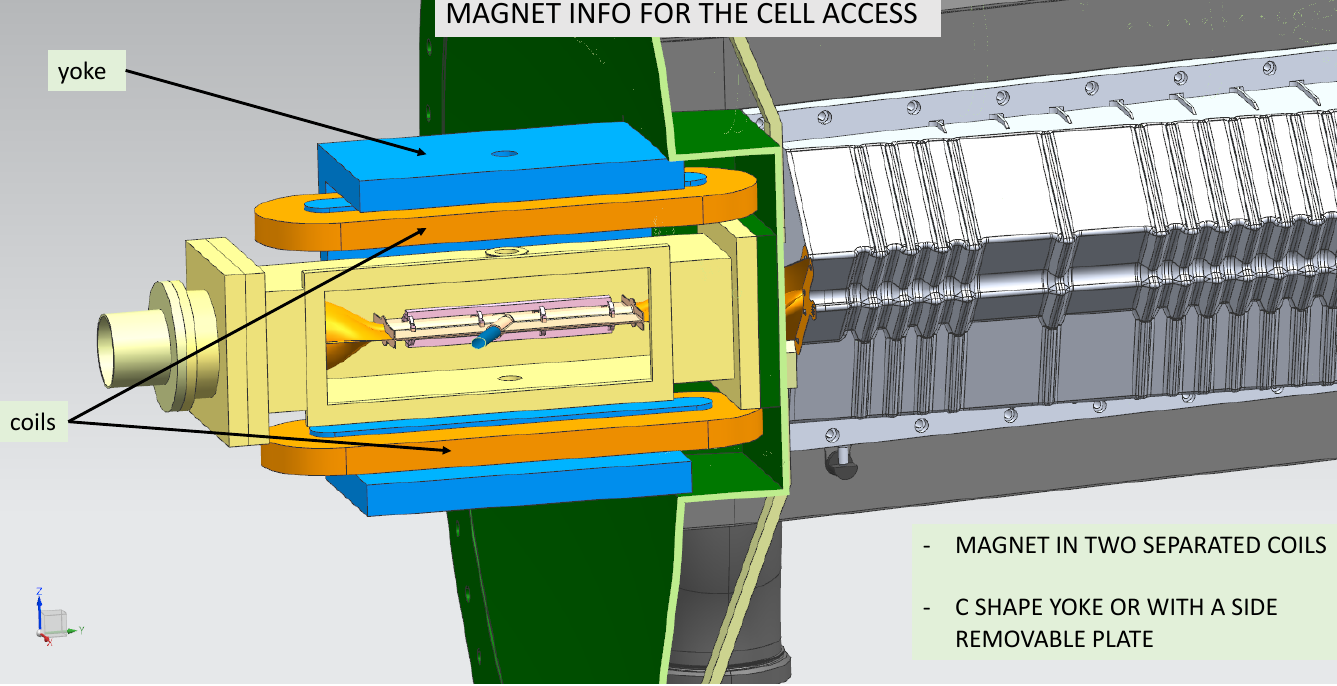}
\caption{\label{draw} A drawing of the PGT with the magnet coils (orange) and the iron return yoke (blue) enclosing the storage cell. The VELO vessel and detector box are shown in green and grey, respectively.}
\end{figure}

The LHCspin experimental setup is in the R$\&$D phase and calls for the development of a new generation polarised target. The starting point is the setup of the polarised target system developed at the HERMES experiment \cite{herm} and comprises three main components: an Atomic Beam Source (ABS), a Polarised Gas Target (PGT) and a diagnostic system. The ABS consists of a dissociator with a coooled nozzle, a Stern-Gerlach apparatus to focus the wanted hyperfine states, and adiabatic RF-transitions for setting and switching the target polarisation. The ABS injects a beam of polarised hydrogen or deuterium into the PGT, which is located in the LHC primary vacuum. The PGT hosts a T-shaped openable storage cell, sharing the SMOG2 general concept, and a compact superconductive dipole magnet, as shown in Fig.~\ref{draw}. The magnet generates a 300 mT static transverse field with a homogeneity of 10 \%, suitable also to avoid beam-induced depolarisation \cite{bid}. Studies for the inner coating of the storage cell are currently ongoing with the aim of producing a surface that minimises the molecular recombination rate as well as the secondary electron yield.

With the use of the SMOG2 system during the LHC Run 3 (2022-2024), the first data usable for studying the mutual target-beam interactions will be available providing a fundamental playground for the R$\&$D of LHCspin. For instance, keeping the PGT in the same location of the SMOG2 cell, the largest kinematic acceptance is achievable. Here, the track reconstruction efficiency remains high with no sensible degradation with respect to the nominal efficiencies reached in the pp collisions at LHCb. New algorithms are currently being developed for the Run 3 fixed-target reconstruction and are expected to sensibly improve the current performance as well as to enable to record LHCspin data in parallel with pp collisions \cite{eff}. The diagnostic system continuously analyses gas samples drawn from the PGT and comprises a target gas analyser to detect the molecular fraction, and thus the degree of dissociation, and a Breit-Rabi polarimeter to measure the relative population of the injected hyperfine states. An instantaneous luminosity of $O(10^{32})$ cm$^{-2}$s$^{-1}$ is foreseen for fixed-target $pH$ collisions in Run 4, with a further factor 3 - 5 increase for the high-luminosity LHC phase from Run 5 (2032).

The fixed-target physics program at LHC has been greatly enhanced with the recent installation of the SMOG2 gas storage cell at LHCb. LHCspin is its natural evolution and aims at installing a polarised gas target to bring spin physics at LHC for the first time, opening a whole new range of exploration. With strong interest and support from the international theoretical community, LHCspin is a unique opportunity to advance our knowledge on several unexplored QCD areas, complementing both existing facilities and the future Electron-Ion Collider \cite{eic}.


\end{document}